\documentclass{tlp}

\makeatletter%
\@mparswitchfalse%
\makeatother%
\reversemarginpar%

\usepackage[utf8]{inputenc}
\usepackage[english]{babel}
\usepackage[table,xcdraw,dvipsnames]{xcolor}
\usepackage{xspace}
\usepackage{multirow}
\usepackage{amsmath,amssymb,stmaryrd}
\usepackage{algorithm}
\usepackage{algorithmicx}
\usepackage[noend]{algpseudocode}
\usepackage{caption}
\usepackage{hyperref}
\hypersetup{%
  colorlinks=true,
  citecolor=blue,
  linkcolor=blue,
  urlcolor=blue,
  filecolor=blue,
  pdftitle={%
    Exploiting Multiple Abstract Call Patterns for Optimizing Run-Time Checks%
  },
  pdfauthor={%
    Daniela Ferreiro,
    Daniel Jurjo-Rivas,
    Marco Ciccal\`{e},
    Jose F.~Morales,
    Pedro L\'{o}pez-Garc\'{i}a,
    Manuel V.~Hermenegildo%
  },
  pdfsubject={Computer Science},
  pdfkeywords={%
    Abstract Interpretation,
    Assertions,
    (Constraint) Logic Programming,
    Verification,
    Run-Time Checking%
  },
  bookmarksopen=true,
  bookmarksnumbered=true
}
\usepackage{url}
\usepackage[capitalize,noabbrev]{cleveref}

\usepackage{tikz}
\usetikzlibrary{shapes.geometric}
\usetikzlibrary{matrix}
\usetikzlibrary{shapes.gates.logic.US}
\usetikzlibrary{circuits}
\usetikzlibrary{positioning,shapes,arrows,fit,shadows,calc,matrix}
\usetikzlibrary{backgrounds}
\usetikzlibrary{arrows}

\usepackage{listings}
\usepackage{textcomp}
\definecolor{ciaoframe}     {rgb}{  0,    0,  0.3}
\definecolor{ciaostring}    {rgb}{0.6, 0.46, 0.33}
\definecolor{ciaooperators} {rgb}{0.1, 0.15,  0.6}
\definecolor{ciaokeywords}  {rgb}{0.1, 0.15,  0.6}
\definecolor{ciaoassertions}{rgb}{0.1, 0.15,  0.6}
\definecolor{ciaotrust}     {rgb}{200, 130,     0}
\definecolor{ciaocheck}     {rgb}{0.1, 0.2,   0.8}
\definecolor{ciaochecked}   {rgb}{0.2, 0.34,  0.1}
\definecolor{ciaotrue}      {rgb}{0.2, 0.34,  0.1}
\definecolor{ciaofalse}     {rgb}{0.6,  0.0, 0.09}
\definecolor{ciaoprops}     {rgb}{0.1,  0.2,  0.8}
\definecolor{ciaocomment}   {rgb}{0.5,  0.5,  0.5}
\lstdefinelanguage{Ciao}{%
  columns=fullflexible, %
  language=Prolog,
  xleftmargin=4mm,
  breaklines=true,
  tabsize=4,
  breaklines=true,breakatwhitespace=true,
  basicstyle=\ttfamily,
  showlines=true,
  showspaces=false,
  showtabs=false,
  mathescape=true,
  escapeinside={(*}{*)},
  commentstyle=\color{ciaocomment},
  stringstyle=\color{ciaostring},
  showstringspaces=false,
  deletekeywords={true,pp,is,var}, %
  keywordstyle={\color{ciaooperators}\bfseries}, %
  classoffset=1, %
  otherkeywords={=>},
  keywordstyle={\color{ciaokeywords}\bfseries},
  classoffset=2,
  morekeywords={module,use_module,dynamic,export,import,impl_defined,trait,impl},
  keywordstyle={\color{ciaokeywords}\bfseries},
  morekeywords={pred,prop,calls,success,comp,compat,inst,modedef,regtype},
  keywordstyle={\color{ciaoassertions}\bfseries},
  classoffset=4,
  morekeywords={trust,trust_default,entry},
  keywordstyle={\color{ciaotrust}\bfseries},
  classoffset=5,
  morekeywords={check},
  keywordstyle={\color{ciaocheck}\bfseries},
  classoffset=6,
  morekeywords={checked},
  keywordstyle={\color{ciaochecked}\bfseries},
  classoffset=7,
  morekeywords={true},
  keywordstyle={\color{ciaotrue}\bfseries},
  classoffset=8,
  morekeywords={false},
  keywordstyle={\color{ciaofalse}\bfseries},
  classoffset=0
}
\lstdefinelanguage{CiaoLong}{%
  language=Ciao,
  basicstyle=\footnotesize\ttfamily,
  frame=l,
  framesep=5pt,
  rulecolor=\color{black},
  numbers=none,%
  xleftmargin=6pt,xrightmargin=4pt
}
\newcommand{\ciaoinline}[1]{%
  \ifmmode%
    \text{\lstinline[language=Ciao]{#1}}%
  \else%
    \lstinline[language=Ciao]{#1}%
  \fi%
}
\newcommand{\clp}{(C)LP\xspace}
\newcommand{\plai}{PLAI\xspace}
\newcommand{\ciao}{Ciao\xspace}

\newcommand{\lpdoc}{LPdoc\xspace}

\newcommand{\andtrees}{\textsc{and}-trees\xspace}

\newcommand{\ciaochecked}{\texttt{\color{ciaochecked}checked}\xspace}
\newcommand{\ciaofalse}{\texttt{\color{ciaofalse}false}\xspace}
\newcommand{\ciaocheck}{\texttt{\color{ciaocheck}check}\xspace}

\newtheorem{definition}{Definition}[section]
\newtheorem{example}{Example}[section]

\renewcommand{\emptyset}{\ensuremath{\varnothing}}
\newcommand{\quantSep}{\ensuremath{\ldotp\;}}
\newcommand*{\seq}[2][n]{{#2_{1},\allowbreak\ldots,\allowbreak #2_{#1}}}

\newcommand{\setcomp}[2]{\ensuremath{\{#1~|~#2\}}}
\newcommand{\tuple}[1]{\langle #1 \rangle}

\newcommand{\prog}{\ensuremath{\mathcal{P}}}

\newcommand{\vars}[1]{\ensuremath{\mathsf{vars}(#1)}}
\newcommand{\clause}{\ensuremath{\head~\texttt{:-}~\body}}

\newcommand{\entails}{\models}

\newcommand{\cquery}{\ensuremath{Q}}

\newcommand{\emptyGoal}{\square}
\newcommand{\state}[2]{\ensuremath{\langle #1\, |\, #2 \rangle}}
\newcommand{\red}[2]{\ensuremath{#1\leadsto #2}}
\newcommand{\redstar}[2]{\ensuremath{#1\leadsto^{*} #2}}

\newcommand{\answers}[1]{\ensuremath{\mathsf{answers}(#1)}}

\newcommand{\checklit}[1]{\ensuremath{\mathsf{c}(#1)}}
\newcommand{\errorlit}[1]{\ensuremath{\mathsf{e}(#1)}}

\newcommand*{\redA}[3][\A]{#2\leadsto_{#1} #3}

\newcommand{\A}{\ensuremath{\mathcal{A}}}

\newcommand{\pred}{\ensuremath{\mathit{Pred}}}
\newcommand{\pre}{\ensuremath{\mathit{Pre}}}
\newcommand{\post}{\ensuremath{\mathit{Post}}}
\newcommand{\acCall}[2]{\ensuremath{\mathsf{calls}(#1,\allowbreak#2)}}
\newcommand{\acSucc}[3]{\ensuremath{\mathsf{success}(#1,\allowbreak#2,\allowbreak#3)}}

\newcommand{\inlineAssrt}[3]{%
  \ciaoinline{:-}~\ciaoinline{pred}~#1~%
  \ciaoinline{:}~#2~%
  \ciaoinline{=>}~#3%
}

\newcommand{\succs}{\ensuremath{\lambda^s}\xspace}
\newcommand{\call}{\ensuremath{\lambda^c}\xspace}
\newcommand{\aprime}[1]{\ensuremath{\lambda^p_{#1}\xspace}}
\newcommand{\entry}{\ensuremath{\lambda^{en}}\xspace}
\newcommand{\exit}{\ensuremath{\lambda^{ex}}\xspace}
\newcommand{\proj}{\ensuremath{\lambda^{pr}}\xspace}
\newcommand{\anGraph}{\ensuremath{\mathcal{G}}}
\newcommand{\augment}{\ensuremath{\mathsf{augment}\xspace}}

\newcommand{\project}{\ensuremath{\mathsf{project}\xspace}}
\newcommand{\extend}{\ensuremath{\mathsf{extend}\xspace}}
\newcommand{\calltoentry}{\ensuremath{\mathsf{callToEntry}\xspace}}
\newcommand{\exittoprime}{\ensuremath{\mathsf{exitToPrime}\xspace}}
\newcommand{\entrytoexit}{\ensuremath{\mathsf{entryToExit}\xspace}}
\newcommand{\head}{\ensuremath{H}}
\newcommand{\goal}{\ensuremath{L}}
\newcommand{\body}{\ensuremath{B}}%
\newcommand{\abscall}{\ensuremath{\mathsf{absAssrtProjs}\xspace}}
\newcommand{\absprime}{\ensuremath{\mathsf{absAssrtPrimes}\xspace}}
\newcommand{\abscalls}[2]{\ensuremath{\abscall(#1,#2)}} %
\newcommand{\absprimes}[3]{\ensuremath{\absprime(#1,#2,#3)}}

\newcommand{\ct}{\textbf{ct}\xspace}
\newcommand{\ctrt}{\textbf{ctrt}\xspace}
\newcommand{\ctrtf}{\textbf{ctrt\_f}\xspace}
\newcommand{\ctrtfetsh}{\textbf{ctrt\_f\_vers\_etsh}\xspace}
\newcommand{\ctrtfshet}{\textbf{ctrt\_f\_vers\_shet}\xspace}
\newcommand{\shfrexp}{\textit{Sharing-Freeness}\xspace}
\newcommand{\etermsexp}{\textit{Eterms}\xspace}

\begin{document}
\lefttitle{D.~Ferreiro et al.}
\jnlPage{1}{8}
\jnlDoiYr{2021}
\doival{10.1017/xxxxx}
\title[%
  Exploiting Multiple Abstract Call Patterns for Optimizing Run-Time
  Checks%
]{%
  Exploiting Multiple Abstract Call Patterns for Optimizing Run-Time
  Checks%
  \thanks{%
    Partially funded by MICIU project CEX2024-001471-M
    \emph{Mar\'{\i}a de Maeztu}
    and by the European Union MSCA GA 101154447 NEAT.
    We would also like to thank the anonymous reviewers for their very
    useful and constructive feedback.%
  }%
}
\begin{authgrp}
  \author{%
    \href{https://orcid.org/0009-0002-1072-8989}{DANIELA FERREIRO},
    \href{https://orcid.org/0000-0001-6215-1080}{DANIEL JURJO-RIVAS},
    \\
    \href{https://orcid.org/0009-0000-8821-0587}{MARCO CICCALÈ} and
    \href{https://orcid.org/0000-0001-9782-8135}{JOSE~F.~MORALES}\\[.6mm]
  }
  \affiliation{%
    Universidad Polit\'{e}cnica de Madrid (UPM),
    IMDEA Software Institute, Madrid, Spain\\[.6mm]
    {\rm(}e-mails:
    {\fontfamily{cmtt}\selectfont\upshape
       \href{mailto:d.ferreiro@alumnos.upm.es}{d.ferreiro@alumnos.upm.es}{\rm\it, }%
       \href{mailto:daniela.ferreiro@imdea.org}{daniela.ferreiro@imdea.org}{\rm\it, }%
       \href{mailto:daniel.jurjo@alumnos.upm.es}{daniel.jurjo@alumnos.upm.es}{\rm\it, }%
       \href{mailto:daniel.jurjo@imdea.org}{daniel.jurjo@imdea.org}{\rm\it, }%
       \href{mailto:m.ciccale@alumnos.upm.es}{m.ciccale@alumnos.upm.es}{\rm\it, }%
       \href{mailto:marco.ciccale@imdea.org}{marco.ciccale@imdea.org}{\rm\it, }%
       \href{mailto:josefrancisco.morales@upm.es}{josefrancisco.morales@upm.es}{\rm\it, }%
       \href{mailto:josef.morales@imdea.org}{josef.morales@imdea.org}}%
     {\rm)}\\[1mm]}
   \author{%
     \href{https://orcid.org/0000-0002-1092-2071}{PEDRO LÓPEZ-GARCÍA}\\[.6mm]}
   \affiliation{%
     Spanish Council for Scientific Research,
     IMDEA Software Institute, Madrid, Spain\\[.6mm]
    {\rm(}e-mails:
    {\fontfamily{cmtt}\selectfont\upshape\href{mailto:pedro.lopez@csic.es}{pedro.lopez@csic.es}{\rm\it, }%
       \href{mailto:pedro.lopez@imdea.org}{pedro.lopez@imdea.org}}%
     {\rm)}\\[1mm]}
   \author{%
     \href{https://orcid.org/0000-0002-7583-323X}{MANUEL V.~HERMENEGILDO}\\[.6mm]}
   \affiliation{%
     Universidad Polit\'{e}cnica de Madrid (UPM),
     IMDEA Software Institute, Madrid, Spain\\[.6mm]
    {\rm(}e-mails:
    {\fontfamily{cmtt}\selectfont\upshape\href{mailto:manuel.hermenegildo@upm.es}{manuel.hermenegildo@upm.es}{\rm\it, }%
       \href{mailto:manuel.hermenegildo@imdea.org}{manuel.hermenegildo@imdea.org}}%
     {\rm)}\\[-4mm]}
\end{authgrp}

\history{\sub{xx xx xxxx;} \rev{xx xx xxxx;} \acc{xx xx xxxx}}

\maketitle       %

\begin{abstract}

  In strongly-typed languages, types are verified at compile time,
  while dynamically typed languages, such as Prolog, perform type
  consistency checks entirely at run-time.  Extending dynamic
  languages with assertions allows expressing both classical types and
  more general properties, providing high expressiveness, but at the
  cost of run-time overhead.  Abstract interpretation allows safely
  approximating such program properties at compile time, which has
  been used to reduce the number %
  of properties that require
  run-time checks, while still reporting
  unverified properties that can guide further static analyses,
  testing, or domain refinement.  In this work, we first study how
  to selectively
  integrate the run-time semantics of assertion properties into a
  multivariant, top-down, goal-directed abstract interpretation
  algorithm. We then show how multiple inferred calling patterns can
  be exploited to reduce the number of properties that must be checked
  at run-time, thus minimizing the overhead.  Finally, we report on an
  implementation of our approach in the Ciao system and provide
  performance results showing improvements over previously reported
  techniques.

%
%
%
%
%
%
%
%
%
%
%
%
%
%
%
%
%
%
%
%
%
%

%
%
%
%
%
%
%
%
%
%
%
%
%
%
%
%
%
%
%
%
%
%
%
%
\end{abstract}

\section{Introduction}

Detecting incorrect program behaviors %
during the compilation phase is an important and complex part of the
software development cycle.
With the advent of machine-generated code, and as more non-experts are
able to write complex software, ensuring that
semi-automatically generated programs behave as expected has become
increasingly important.
To aid in this process, a number of tools have been developed
to compare actual program behavior against expectations, such as code
analyzers/verifiers and run-time verification frameworks.
These tools rely on language-level constructs to describe expected
program behavior.
Approaches that fall into this category include
assertion-based frameworks used in (Constraint) Logic Programming
((C)LP)~\citep{%
  DNM88-short,%
  assert-lang-ws-short-alt,%
  aadebug97-informal-short,%
  BDM97-short,%
  prog-glob-an-short,%
  assrt-theoret-framework-lopstr99-short,%
  DBLP:conf/discipl/Lai00,%
  ciaopp-sas03-journal-scp-short,%
  testchecks-iclp09-short,%
  Hanus17LOPSTR-short%
},
soft/gradual typing
approaches in functional
programming~\citep{
  cartwright91:soft_typing-short,%
  DBLP:conf/icfp/FindlerF02-short,%
  TypedSchemeF08-short,%
  DBLP:journals/toplas/DimoulasF11,%
  DBLP:conf/popl/RastogiSFBV15-short,%
  DBLP:conf/ecoop/TakikawaFDFFTF15-short,%
  DBLP:conf/popl/TakikawaFGNVF16-short,%
  DBLP:journals/pacmpl/VazouTH18} 
and
contract-based extensions in object-oriented
programming~\citep{%
  lamport99:types_spec_lang,%
  DBLP:journals/fac/LeavensLM07,%
  clousot-2010-short%
}.
These approaches often involve a certain degree of run-time testing,
especially for non-trivial properties.
However, such testing can incur significant performance
overhead~\citep{testchecks-iclp09-short}, even
when performing simple type checks between typed and untyped
parts of programs~\citep{DBLP:conf/popl/RastogiSFBV15-short,%
  DBLP:conf/popl/TakikawaFGNVF16-short}.

Some proposals have been made to reduce the run-time
overhead of assertion checking by optimizing the run-time checking
mechanisms themselves, at the expense of increased memory
consumption~\citep{rv2014-short,cached-rtchecks-iclp2015-short}.
Repeated checks on immutable recursive data structures are converted
from execution-time overhead into increased memory use via caching
and/or tabling techniques.
Despite these advances, full 
run-time checking often remains impractically expensive, especially for
complex properties such as deep data structure tests. This reduces the
attractiveness of run-time checking to programmers, who may allow
sporadic checking of very simple conditions but tend to disable
run-time checking for more complex properties.
Motivated by this problem, assertion-based frameworks have been
proposed in which static analysis is
used
to reduce the number of program points where run-time checks are
required, or to rule out incorrect program
behaviors~\citep{assert-lang-ws-short-alt,aadebug97-informal-short,prog-glob-an-short}.
A number of practical \emph{assertion checking modes} have
been
studied, representing different trade-offs between code annotation
depth, execution-time slowdown, and program behavior safety guarantees
\citep{optchk-journal-scp-short}.
The latter proposed a method to modify the semantics
inferred by the analyzer
via \emph{program transformations} to capture executions with active
run-time checking. 
While this technique had limitations, it showed the potential 
to increase the number of checks verified at compile time.

In this work, we take a complementary approach by incorporating
the run-time checking semantics directly into the
analyzer, but making it optional, 
allowing abstract executions to more closely match their concrete
counterparts. By making the presence or absence of run-time checks
part of the analyzed semantics, the analyzer can reason precisely
about program executions under different assertion checking
configurations.
We also study the effect of 
multivariant analysis 
in this setting, which  
makes it possible to distinguish between different calling contexts and
different versions of the same predicate or procedure, depending on
the properties known to hold and on whether run-time checks are
required.
Multivariant analysis is particularly well suited to \clp, where predicates
are naturally invoked under different modes of use. If such modes
are not taken into account, the analysis is forced to merge distinct
execution contexts, %
often leading to a
loss of precision and overly conservative decisions regarding run-time
checking. By explicitly representing different analysis versions
corresponding to different calling modes and run-time checking
configurations, we study how making these versions explicit can
improve analysis precision and run-time checking efficiency.
Finally, beyond measuring how many assertions are completely verified, 
as is common practice, we also evaluate whether individual
properties within an assertion are checked.
Note that assertions often consist of multiple properties of varying
complexity: 
for example, in a formula such as 
$(\mathit{simple\_property}(X),\allowbreak \mathit{complex\_property}(Y),\allowbreak
\mathit{less\_complex\_property}(Z))$, some properties (such as
$\mathit{simple\_property(X)}$) may be fully verified statically, 
others ($\mathit{complex\_property}(Y)$) may never be discharged, and
yet others ($\mathit{less\_complex\_property}(Z)$) may be verified only under
certain calling contexts. 
We aim to capture this finer-grained information, allowing
us to quantify how many properties of each assertion are verified,
under which conditions, and how many remaining properties require
run-time checking, with special attention to multivariance. 
\section{Preliminaries and notation}
\label{sec:prel}
Variables start with a capital letter.
The set of terms is inductively defined as follows: (1) variables are
terms (2) if $f$ is an $n$-ary function symbol and $\seq{t}$ are
terms, then $f(\seq{t})$ is a term.
An \emph{atom} has the form $p(\seq{t})$ where $p$ is an $n$-ary
predicate symbol, and $\seq{t}$ are terms.
A \emph{constraint} is a conjunction of expressions built from
predefined predicates whose arguments are constructed using predefined
functions and variables, \emph{e.g.}, $X-Y > \mathit{abs}(Z)$.
A \emph{literal} is either an atom or a constraint.
We denote the set of variables of a literal $L$ by $\vars{L}$.
\emph{Negation} is encoded as finite failure, supported through a program
expansion.
A \emph{goal} is a finite sequence of literals.
A \emph{rule} has the form $\clause$ where $\head$, the \emph{head},
is an atom and $\body$, the \emph{body}, is a possibly empty finite
sequence of literals.
A \emph{predicate} is a set of rules with the same head.
We refer to a predicate by either its function symbol and arity
$p/n$, or its head $p(\seq{t})$.
A \emph{constraint logic program}, or \emph{program}, is a finite set
of rules.
For simplicity, we assume a single program $\prog$ in which every rule
has distinct variables in its head;
and a class of constraints---over a given constraint domain---for
which projection and a complete solver exist.
\paragraph{\bfseries Operational semantics.}
The operational semantics, adapted from that
of~\citet{jaffar98-clp-sem}, is given in terms of
\emph{derivations} from goals.
Derivations are sequences of \emph{reductions} between \emph{states}.
A state $\state{G}{\theta}$ consists of a goal $G$, and the current
constraint $\theta$.
We denote sequence concatenation by (::).
We use $\red{S}{S'}$ to indicate that a reduction step can be applied
to state $S$ to obtain state $S'$.
And $\redstar{S}{S'}$ to indicate that there exists a sequence of
reduction steps from $S$ to $S'$.
\Cref{fig:op-sem} depicts the reduction rules.
\begin{figure}
  \begin{align}
    \state{\theta' :: G}{\theta} \leadsto
    &\; \state{G}{\theta \wedge \theta'}
    &&\qquad\mathit{if~}
       (\theta \wedge \theta') \not\entails \mathit{false}
    &\qquad(\textsc{Constr})\tag*{}\label{rule:constr}\\
    \state{\head :: G}{\theta} \leadsto
    &\; \state{\body :: G}{\theta}
    &&\qquad\mathit{if~} \exists (\clause) \in \prog
    &\qquad(\textsc{Atom})\tag*{}\label{rule:atom}
  \end{align}
  \caption{Reduction rules for the operational semantics of \clp programs.}
  \label{fig:op-sem}
\end{figure}
Intuitively, satisfiable constraints are added to the current
constraint (\hyperref[rule:constr]{\textsc{Constr}}),
and atoms are reduced by unfolding them into the bodies of matching
program rules (\hyperref[rule:atom]{\textsc{Atom}}).

Given a literal $L$ and reduction steps $\redstar{S}{S'}$ from state
$S = \state{L::G}{\theta}$ to state $S' = \state{G}{\theta'}$, we
refer to $S$ as a \emph{call state} for $L$, and to $S'$ as a
\emph{success state} for $L$.

A \emph{query} is a pair $(L,\theta)$, where $L$ is a literal and
$\theta$ is a constraint for which the \clp system starts a
computation from state $\state{L}{\theta}$.
A finite derivation from a query $\cquery=(L,\theta)$ is
\emph{finished} if the last state in the derivation cannot be reduced.
Such a finished derivation is \emph{successful} if the last state is
of the form $\state{\emptyGoal}{\theta'}$, where $\emptyGoal$ denotes
the empty goal sequence, and the projection of the constraint
$\theta'$ onto the variables $\vars{L}$ is an \emph{answer} to
$\cquery$.
Conversely, a derivation is \emph{failed} if the last state is not of
the form $\state{\emptyGoal}{\theta}$.
We denote by $\answers{\cquery}$ the set of answers
to the query $\cquery$.
\paragraph{\bfseries Properties \& property formulas.}
Conditions on the current constraint are stated as \emph{property
  formulas}, which are conjunctions of \emph{properties}.
Properties are predicates, typically defined in the source language,
and thus runnable.
However, not all predicates can be used as properties. They are
generally 
required to be checkable at run time so that they can be effectively
used as run-time checks, but not necessarily decidable at compile
time, where they can be safely
approximated~\citep{prog-glob-an-short,assrt-theoret-framework-lopstr99-short}.
\begin{example}[Properties]\itshape
  \label{ex:props}
  The following are the definitions of the $\mathit{list/2}$,
  $\mathit{tree/2}$ and $\mathit{list\_or\_tree/2}$ properties:
  {\rm
\begin{lstlisting}
% Lists (*~~~~\,~~~~~~~~~~~~~~~~*) % Binary trees (*~~~~\,~~~~~~~~~~~~~~*) % Lists or binary trees
:- prop list/2. (*~~~~~~~~~~~~\,*) :- prop tree/2. (*~~~~~~~~~~~~~~~~~\,*) :- prop list_or_tree/2.
(*\bfseries list*)(_,[]). (*~\;~~~~~~~~~~~~~~~~*)(*\bfseries tree*)(_,void). (*~\,~~~~~~~~~~~~~~~~~~*) (*\bfseries list\_or\_tree*)(P,L) :-
(*\bfseries list*)(P,[X|L]) :- (*~\;~~~~~~~~~~~*)(*\bfseries tree*)(P,tree(X,L,R)) :- (*~\;~~~~~~~~~~~~~*)list(P,L).
    P(X), (*~\;~~~~~~~~~~~~~~~~~~~~~*) P(X), (*~~~~~~~~~~~~~~~~~~~~~~~~*) (*\bfseries list\_or\_tree*)(P,T) :-
    list(P,L). (*~\;~~~~~~~~~~~~~~~~*) tree(P,L), tree(P,R). (*~~\,~~~~~~~~~*) tree(P,T).
\end{lstlisting}
  }
  Note that all definitions are \emph{parametric} on an additional
  property (their first argument), which is then called using
  higher-order notation ({\rm\texttt{P(X)}}).
\end{example}
\paragraph{\bfseries Assertions.}
Assertions are syntactic objects which allow expressing properties of
programs that must be satisfied at certain points of program
execution.
We
recall the relevant parts of the assertion schema
of~\citet{assert-lang-disciplbook-short}. %
\emph{Predicate} (or \emph{pred}) assertions have the following syntax:
\vspace*{-1.5mm} %
\[\inlineAssrt{\pred}{\pre}{\post}\]
where $\pred$ is an atom representing a predicate, and $\pre$ and
$\post$ are property formulas.
They express that all calls to $\pred$ \emph{must} satisfy the
precondition $\pre$, and, if such calls succeed, the
postcondition $\post$ \emph{must} be satisfied.
If there are several \emph{pred} assertions, the
$\pre$ field of at least one of them \emph{must} be satisfied.
To simplify both presentation and use, assertions are normalized into
a set of corresponding \emph{assertion conditions}.

\begin{definition}[Assertion conditions]\itshape
  \label{def:ac}
  Given a predicate $\pred$ and its corresponding set of assertions
  $\{\seq{A}\}$ with
  {\rm$A_i = \inlineAssrt{\pred}{\pre_i}{\post_i}$},
  its set of \emph{assertion conditions} is defined as
  $\{C_0,\seq{C}\}$ with:%
  \[
    C_i =
    \begin{cases}
      \acCall{\pred}{\bigvee_{j=1}^n\, \pre_j} & i = 0\\
      \acSucc{\pred}{\pre_i}{\post_i} & i \in \{1, \ldots, n\}
    \end{cases}\\[1.5mm]
  \]
  $C_0$ encodes the checks that the calls to the predicate represented
  by $\pred$ are within those admissible by the set of assertions.  We
  refer to it as the \emph{calls assertion condition}.
  $\seq{C}$ encode the checks for compliance of the
  success states
  for
  particular sets of calls, and we call them the \emph{success
    assertion conditions}.
\end{definition}

We denote by $\A$ both the set of assertions of the program and,
interchangeably, its associated set of assertion conditions.
And we assume that, if there are no assertions associated with a
predicate $\pred$, its set of assertion conditions is defined as
$\{\acCall{\pred}{\mathit{true}},\allowbreak\acSucc{\pred}{\mathit{true}}{\mathit{true}}\}$;
that is, the most-general assertion/assertion conditions.

\begin{example}[Member program with assertions]\itshape
  \label{ex:member}
  Consider the program in Figure~\ref{code:member}, which defines the predicate
  $\mathit{member(X,S)}$, relating a number $X$ with a structure $S$.
  As specified by its assertion, it succeeds when the number $X$
  occurs at some position in the structure $S$, where $S$ may be a
  \emph{list} or a \emph{binary tree}.
\begin{figure}[t]
\begin{lstlisting}
:- pred member(X,S) : (var(X), list_or_tree(num,S)) => num(X).

(*\bfseries member*)(X,tree(X,L,R)) :- tree(num,L), tree(num,R).
(*\bfseries member*)(X,tree(_,L,_)) :- member(X,L). (*~~~~~~~~~~~~~~~~*)(*\bfseries member*)(X,[X|L]) :- list(num,L).
(*\bfseries member*)(X,tree(_,_,R)) :- member(X,R). (*~~~~~~~~~~~~~~~~*)(*\bfseries member*)(X,[_|L]) :- member(X,L).
\end{lstlisting}
     \caption{An example Prolog program with an assertion capturing its behavior.\label{code:member}}
\end{figure}
\end{example}

\paragraph{\bfseries Run-time assertion checking.}
\begin{figure}
  \begin{align}
    \state{\theta' :: G}{\theta} \leadsto_{\A}
    &\, \state{G}{\theta \wedge \theta'}
    &&\qquad\mathit{if~}
       (\theta \wedge \theta') \not\entails \mathit{false}
    &\qquad(\textsc{Constr}_{\A})\tag*{}\label{rule:constr-assrt}\\
    \state{\head :: G}{\theta} \leadsto_{\A}
    &\, \state{\mathsf{wrap}_{\A}(\head,\body,\theta) :: G}{\theta}
    &&\qquad\mathit{if~} \exists (\clause) \in \prog
    &\qquad(\textsc{Atom}_{\A})\tag*{}\label{rule:atom-assrt}\\
    \state{\checklit{F} :: G}{\theta} \leadsto_{\A}
    &\, \state{G}{\theta}
    &&\qquad\mathit{if~} \mathsf{check}(F,\theta)
    &\qquad(\textsc{OkCheck}_{\A})\tag*{}\label{rule:okCheck-assrt}\\
    \state{\checklit{F} :: G}{\theta} \leadsto_{\A}
    &\, \state{\mathsf{e}(F)}{\theta}
    &&\qquad\mathit{if~} \neg\mathsf{check}(F,\theta)
    &\qquad(\textsc{ErrCheck}_{\A})\tag*{}\label{rule:errCheck-assrt}
  \end{align}
  \begin{align*}
    \mathsf{check}(F,\theta) \triangleq
    &\; \exists \theta' \in \answers{(F,\theta)} \quantSep
       \theta \entails \theta'\\
    \mathsf{wrap}_{\A}(\head,\body,\theta) \triangleq
    &\; \checklit{\pre} :: \body :: \setcomp{%
      \checklit{\post_i}
   }{\acSucc{\head}{\pre_i}{\post_i} \in \A \wedge \mathsf{check}(\pre_i,\theta)      
    },\\[-.5mm]
    &\:\,\mathit{with~} \exists! \acCall{\head}{\pre} \in \A
  \end{align*}
  \caption{Reduction rules for the operational semantics of \clp programs with assertions.}
  \label{fig:op-sem-assrt}
\end{figure}
Run-time assertion checking consists of \emph{dynamically} checking
that the conditions
imposed by
the assertions hold while computing the derivations from a
query.
To this end, we incorporate instrumental checks into the operational
semantics in Figure~\ref{fig:op-sem}.
We extend the set of literals with additional syntactic objects:
\emph{check} literals of the form $\checklit{F}$, where $F$ is a
property formula to be checked at run-time; and \emph{error} literals of
the form $\errorlit{F}$, where $F$ is a property formula that has been
violated.

We use $\redA{S}{S'}$ to indicate run-time checking reductions
\emph{w.r.t.}~a set of assertions $\A$.
\Cref{fig:op-sem-assrt} depicts the reduction rules and some auxiliary
functions.
Intuitively, atoms are still reduced by unfolding them into the bodies
of matching program rules.
However, the body is now preceded by a run-time check for the
predicate's pre-condition, \emph{i.e.}, its \emph{calls} assertion
condition, and followed by a sequence of run-time checks for the
predicate's applicable post-conditions, \emph{i.e.}, its set of
\emph{success} assertion conditions whose pre-condition held at the
atom's call state (\hyperref[rule:atom-assrt]{\textsc{Atom}$_{\A}$}).
Run-time checks are computed by means of the $\mathsf{check}$
function, which determines whether a property formula $F$ holds under
the current constraint $\theta$ by searching for a solution for $F$
that is entailed by $\theta$.
Since run-time checks may fail,
we introduce the notion of
(finite) \emph{erroneous} derivations, whose final state is of the
form $\state{\errorlit{F}}{\theta}$
indicating
that the run-time check of $F$ under
$\theta$ was \emph{false}
(\hyperref[rule:errCheck-assrt]{\textsc{ErrCheck}$_{\A}$}).

While useful, 
run-time assertion checking can be prohibitively expensive in both
time and memory. Consider calling the $\mathit{member/2}$ predicate
in~\cref{ex:member} with the calling mode expressed by
\ciaoinline{:-}~\ciaoinline{pred}~$\mathit{member}(X,S)$~%
\ciaoinline{:}~$(\mathit{var}(X),\mathit{list\_or\_tree}(\mathit{num},S))$.
Then, checking that $S$ is a
list at each recursive step turns the $\mathcal{O}(n)$ algorithm into $\mathcal{O}(n^2)$.

\paragraph{\bfseries Abstract interpretation.}
The main idea behind \emph{abstract interpretation}~\citep{Cousot77-short} is
to interpret the program over a special, abstract domain whose
elements are finite representations of possibly infinite sets of
states in the concrete domain. We 
denote the concrete domain as $D_{\gamma}$, the abstract domain as
$D_{\alpha}$, 
and
the functions that relate sets of
states with
their abstractions as the \emph{abstraction} function $\alpha : D_{\gamma} \rightarrow D_{\alpha}$ and the \emph{concretization} function $\gamma : D_{\alpha} \rightarrow D_{\gamma}$.
The concrete domain is typically a complete lattice with the set
inclusion order which induces an ordering relation in the abstract
domain 
represented by $\sqsubseteq$. Under this relation the abstract domain
is usually a complete lattice and
$(D_{\gamma}, \alpha, D_{\alpha}, \gamma)$ is a Galois
insertion/connection~\citep{Cousot77-short}.

\emph{Top-down} analyses are a family of static analyses that build an
\emph{analysis graph} starting from a series of program \emph{entry
  points}. This approach was first used in analyzers such as MA3 and
Ms~\citep{pracabsin-short}, and matured in the \plai
analyzer~\citep{mcctr-fixpt-short,ai-jlp-short} using an optimized
fixpoint algorithm now also referred to as the \emph{top-down
  algorithm} or \emph{solver}. This algorithm was later applied to the
analysis of CLP/CHCs~\citep{anconsall-acm-short}\footnote{The cited
  paper showed that CLP analysis can in fact be done with the same
  techniques used for LP provided suitable abstractions are provided
  in the abstract domains for %
  the constraint primitives.}
and imperative
programs~\citep{anal-peval-horn-verif-2021-tplp-short,HGScam06-short,decomp-oo-prolog-lopstr07-short,fixpt-javabytecode-bytecode07-short},
and used in analyzers such as GAIA~\citep{LeCharlier94:toplas-short},
the CLP($\cal R$) analyzer~\citep{softpe}, or
Goblint~\citep{seidl3improv,tdSeidlA2I-short}.
The graph constructed by the \plai algorithm during analysis 
is a finite, abstract object whose concretization approximates the
(possibly infinite) set of (possibly infinite) maximal \andtrees of
the concrete semantics.
This follows the overall abstraction scheme of~\cite{bruy91}, but
implementing it with an efficient fixpoint algorithm that uses tabling,
dependency-based acceleration, incrementality, etc.
PLAI separates the abstraction of the structure of the
concrete trees (the paths through the program) from the abstraction of
the
\emph{constraints} at the nodes in those
concrete trees (the program states in those paths). The first
abstraction, $T_{\alpha}$, is typically built-in, as an abstract
domain of \emph{analysis graphs}.
The framework is \emph{parametric} on a second abstract domain,
$D_{\alpha}$, whose elements appear as labels in the nodes of the
analysis graph. Each node of the analysis graph is of the form
$\langle \lambda^c,\head,\lambda^s \rangle$ with $\head$ a call
pattern for a predicate of the program and $\lambda^c$ and $\lambda^s$
the abstractions of the call and success states.
\paragraph{\bfseries Multivariance and path-sensitivity.}
The abstract semantics computed by \plai is %
\emph{multivariant}.
That is, for a single program point, it computes separate path
information for each of its \emph{calling contexts} (also referred to
as \emph{variants}, \emph{versions} or \emph{modes}).
In the analysis graph, this is represented by multiple nodes
corresponding to the same program point, each abstracting a different
calling mode.
\medskip

\begin{example}[Analysis]\itshape
  Consider the simple abstract domain depicted in \cref{fig:absdom}.
  \Cref{fig:an-graph} shows a (partial) analysis graph for the program
  in \cref{fig:p}.
  Notice that the success abstraction of $\mathit{member}(X,L)$ becomes the
  call abstraction of $\mathit{member}(X,T)$.
  Notice also how two different versions for $\mathit{member/2}$ do appear,
  with (abstract) call patterns $\{X/\top,L/\mathit{list},T/\mathit{tree}\}$ and
  $\{X/\mathit{num},L/\mathit{list},T/\mathit{tree}\}$, respectively.
\end{example}
\begin{figure}
  \hspace*{-2.2cm}\begin{tikzpicture}
    \node (1)  {$p(L,T)$} ;
    \node (1c) [left=.5mm of 1]  {\footnotesize$\{L/\mathit{list},T/\mathit{tree}\}$} ;
    \node (1s) [right=.5mm of 1] {\footnotesize$\{L/\mathit{list},T/\mathit{tree}\}$} ;
    \node (2) [below=1mm of 1] {$p(L,T)$} ;
    \node (aux) [below of = 2] {};
    \node (3) [left = 1.6cm of aux] {$\mathit{member}(X,L)$} ;
    \node (3c) [text width=1cm, left = 2cm of 3] {\footnotesize$\{X/\top,L/\mathit{list},T/\mathit{tree}\}$} ;
    \node (3s) [text width=1cm, left = 11mm of 3c, below of = 2] {\footnotesize$\{X/\mathit{num},L/\mathit{list},T/\mathit{tree}\}$} ;
    \node (4) [right = 1.6cm of aux] {$\mathit{member}(X,T)$} ;
    \node (4s) [text width=1cm, right = -.5mm of 4] {\footnotesize$\{X/\mathit{num},L/\mathit{list},T/\mathit{tree}\}$} ;
    \node (5) [below = 1mm of 3] {$\mathit{member}(V_1,V_2)$};
    \node (6) [below of = 5] {};
    \node (7) [left of = 6] {};
    \node (8) [right of = 6] {};
    \node (9) [below = 1mm of 4] {$\mathit{member}(V_3,V_4)$};
    \node (10) [below of = 9] {};
    \node (11) [left of = 10] {};
    \node (12) [right of = 10] {};
    \draw[-] (2) edge (3);
    \draw[-] (2) edge (4);
    \draw[loosely dashed] (5) edge (6);
    \draw[loosely dashed] (5) edge (7);
    \draw[loosely dashed] (5) edge (8);
    \draw[loosely dashed] (9) edge (10);
    \draw[loosely dashed] (9) edge (11);
    \draw[loosely dashed] (9) edge (12);
  \end{tikzpicture}
  \caption{Analysis graph of the $\mathit{p/2}$ predicate in \cref{fig:p} with the abstract domain in \cref{fig:absdom}.} 
  \label{fig:an-graph}
  \medskip
  \begin{minipage}[b]{0.5\textwidth}
\begin{lstlisting}
(*\bfseries p*)(L,T) :-
    member(X,L),
    member(X,T).
\end{lstlisting}
    \smallskip
    \caption{Program with different calling modes.}
    \label{fig:p}
  \end{minipage}
  \begin{minipage}[b]{0.48\textwidth}
    \centering
    \begin{tikzpicture}[yscale=.8]
      \node (top)  at (   0,   1.5) {$\top$};
      \node (list)  at ( 1.5, .75) {$\mathit{list}$};
      \node (num)   at (   0, .75) {$\mathit{num}$};
      \node (tree) at (-1.5, .75) {$\mathit{tree}$};
      \node (bot)  at (   0,   0) {$\bot$};
      \draw
      (top)  -- (list)
      (top)  -- (num)
      (top)  -- (tree)
      (list)  -- (bot)
      (num)  -- (bot)
      (tree)  -- (bot);
    \end{tikzpicture}
    \caption{Abstract domain lattice.}
    \label{fig:absdom}
  \end{minipage}
\end{figure}

\paragraph{\bfseries Compile-time assertion verification \&
  simplification.}
Compile-time assertion verification consists of \emph{statically}
checking that the conditions imposed by the assertions hold for
\emph{all} possible derivations of a program from a set of queries.
This verification can be performed using the analysis graph computed
by \plai, which is an over-approximation of the program's semantics
for all such derivations.
However, since \plai is multivariant, a single program point $\head$
may correspond to multiple nodes in the analysis graph:
\(
  \tuple{\lambda^c_1,\head,\lambda^s_1},%
  \ldots,%
  \tuple{\lambda^c_n,\head,\lambda^s_n},
\)
with each node over-approximating the semantics of that point for
\emph{some} derivations.
Consequently, the over-approximation of the semantics of a program
point for \emph{all} derivations is the combination (\emph{lub}) of
all corresponding nodes in the analysis graph:
\(
  \tuple{\sqcup\{\lambda^c_1,\ldots,\lambda^c_n\},%
  \head,%
  \sqcup\{\lambda^s_1,\ldots,\lambda^s_n\}}.
\)
This can then be used for \emph{simplifying} (as much as possible)
the property formulas in the corresponding assertion conditions.
That is, given a program point and its abstraction, properties
\emph{implied} by the abstraction simplify to $\mathit{true}$,
properties \emph{incompatible} with it simplify to $\mathit{false}$,
while the remaining ones remain unchanged.
This approach has the advantage that, even if a property formula
cannot be proved in its entirety---given the undecidable nature of
(non-trivial) semantic properties---it may be possible to discharge
some parts of it.
The verification results are reported as changes in the status and
transformations of the assertions.
An assertion is \ciaochecked if both property formulas are simplified
to $\mathit{true}$, and \ciaofalse if some property formula is simplified
to $\mathit{false}$.
If some property formula cannot be fully proved, the assertion is \ciaocheck,
and its property formulas are replaced with their simplified versions.

From this point on, we assume that the program $\prog$, the set of
assertions $\A$ and the analysis graph $\anGraph$ are implicit
variables.

\section{Assertion checking with abstract run-time checking semantics}
\label{sec:assrt:rt}

\noindent The operational semantics of \clp with run-time assertion
checking (cf.~\cref{fig:op-sem-assrt}) differs from the default \clp
semantics (cf.~\cref{fig:op-sem}) in that the latter can be seen as an
over-approximation of the former.  Intuitively, run-time assertion
checking may cause certain program points to become unreachable due to
failing checks, whereas they would be reachable under the default
semantics. As a result, for a given program $\prog$, the set of
\andtrees generated under the default \clp semantics is a superset of
those generated when run-time assertion checking is enabled.
Thus, abstract interpretation analyses based on the default
semantics are correct but may suffer from a loss of precision, since they
abstract a set of \andtrees that over-approximates the actual execution
space. In contrast, analyzing programs under run-time checking
semantics allows us to exploit the additional information provided by
assertions.
Concretely, when entering a predicate definition, we may assume that
the corresponding calls assertion condition holds, and thus enrich the
call abstraction $\call$ with this information.
Similarly, upon predicate exit, we can assume that the relevant
success assertion conditions hold and use them to refine the resulting
success abstraction $\succs$.

\cite{optchk-journal-scp-short} propose a program transformation
based on the introduction of additional link predicates with assertions
that enriches the analysis with the run-time checking semantics.
In contrast, in this work we focus on integrating run-time checking
semantics directly into the top-down abstract interpretation framework
without transforming the original program.
However, since the information in the assertions is now included in
the abstractions, those abstractions would trivially satisfy the
assertions they are intended to check, turning the analysis results
unsuitable for assertion checking.
To address this, we generate alternative call and success abstractions
at those program points where the assertions' information causes a
gain in precision, and use these during the (abstract) assertion
checking phase to ensure correct behavior.
We now describe this technique in detail.
\paragraph{\bfseries Assertion-based abstraction refinement.}
Assume a literal \goal~in the body of some clause in
the program, and a rule $\clause$ \emph{s.t.}~\head~unifies with \goal.
Assume also that the
constraint affecting \goal~at the
time of this call is approximated by the abstraction \call
such that $\vars{\goal}\subseteq\mathsf{dom}(\call)$ and
$\vars{\call}\cap(\vars{\head}\cup\vars{\body})=\emptyset$.
The success (exit state) of \goal~after executing the rule above is
represented by the abstraction \succs~given by:
\begin{align*}
  \succs  &= \extend(\call,\goal,\aprime{rt}) \\
  \aprime{rt}&=\absprimes{\proj_{rt}}{\aprime{}}{\goal} \\
  \aprime{} &= \exittoprime(\project(\vars{\head}, \exit), \head, \goal) \\
  \exit  &= \entrytoexit(\entry, \head, \body) \\
  \entry &= \augment(\vars{\body}\backslash\vars{\head}, \calltoentry(\proj_{rt}, \goal, \head)) \\
  \proj_{rt}&=\abscalls{\proj}{\goal} \\
  \proj  &= \project(\vars{\goal}, \call) 
\end{align*}

The analyzer operates as follows:
(\textbf{1}) First, it projects the call abstraction (\call) over the
variables in the literal (\goal).
(\textbf{2}) The computed \proj assertion is then enriched by
incorporating the \emph{glb} ($\sqcap$) of the assertion
pre-conditions for the predicate of \goal.
This is achieved by invoking the \abscall{}
(\cref{alg:abscalls}).
If the resulting abstraction is more precise than the initial \proj,
an alternative call abstraction is stored for use during the assertion
checking phase (\cref{alg:abscalls},
lines~\ref{alg:abscalls:store:begin}--\ref{alg:abscalls:store:end}).
(\textbf{3}) Next, it performs abstract unification using the
\calltoentry~function.
(\textbf{4}) The resulting abstraction is then augmented with any variables in the body
not present in the head.
(\textbf{5}) The clause body is then traversed, and
each literal is analyzed using the \entrytoexit~function, producing an
exit abstraction (\exit).
(\textbf{6}) The abstract unification from step (2) is reverted via
\exittoprime, yielding a prime abstraction (\aprime{}).
(\textbf{7}) This abstraction is then enriched with the applicable
assertion postconditions by computing
\absprime{}
(\cref{alg:absprimes}).
As in step (2), if the abstraction obtained is more precise than the
initial one, an alternative success abstraction is stored
(\cref{alg:absprimes},
lines~\ref{alg:absprimes:store:begin}--\ref{alg:absprimes:store:end}).
(\textbf{8}) Finally, this abstraction is incorporated into the
original call, returning the success abstraction (\succs) through the
\extend~function.

As some final remarks, if no predicate head can be unified with the
literal under analysis, a bottom abstraction ($\bot$) is returned
(representing that the exit state is unreachable). If several clauses
are available, all of them are analyzed, and a collection of
\emph{prime} abstractions $\aprime{1},\ldots,\aprime{m}$ is obtained, one
abstraction per clause, where $m$ is the number of clauses. Then, the
success abstraction is computed as
$\succs = \extend(\call,\sqcup\{\aprime{1},\ldots,\aprime{m}\})$ by means
of the \emph{lub} of the collection of
abstractions (other operators, including disjunction and widenings,
are possible).
\begin{figure}
\begin{minipage}[t]{.4\textwidth}
  \begin{algorithm}[H]
    \caption{\abscall~function.}
    \label{alg:abscalls}
    \scriptsize
    \begin{algorithmic}[1]
      \algrenewcommand\alglinenumber[1]{\color{gray}\scriptsize #1}
      \Function{\abscall}{\proj,\goal}
        \State $\mathcal{C} \gets \setcomp{%
            \alpha(F_c)
          }{%
            \acCall{\goal}{F_c} \in \A
          }$
        \State $\proj_{rt} \gets \sqcap(\mathcal{C} \cup \{\proj\})$
        \If{$\proj_{rt} \sqsubseteq \proj $}\label{alg:abscalls:store:begin}
          \State $\mathsf{storeAltCall}(\goal,\proj)$
        \EndIf
        \State \Return  $\proj_{rt}$
      \EndFunction\label{alg:abscalls:store:end}
    \end{algorithmic}
  \end{algorithm}
\end{minipage}
\hfill
\begin{minipage}[t]{.55\textwidth}
  \begin{algorithm}[H]
    \caption{\absprime~function.}
    \label{alg:absprimes}
    \scriptsize
    \begin{algorithmic}[1]
      \algrenewcommand\alglinenumber[1]{\color{gray}\scriptsize #1}
        \Function{\absprime}{\proj,\aprime{},\goal}
          \State $\mathcal{S} \gets \setcomp{\alpha(F_s)}{\acSucc{\goal}{F_c}{F_s} \in \A
            \wedge \proj \sqsubseteq \alpha(F_c)}
            $
        \State $\aprime{rt} \gets \sqcap(\mathcal{S} \cup \{\aprime{}\})$
        \If{$\aprime{rt} \sqsubseteq \aprime{} $} \label{alg:absprimes:store:begin}
        \State $\mathsf{storeAltSucc}(\goal,\proj,\aprime{})$
        \EndIf
        \State \Return  $\aprime{rt}$
        \EndFunction \label{alg:absprimes:store:end}
    \end{algorithmic}
  \end{algorithm}
\end{minipage}
\end{figure}
\section{Exploiting versions for more optimization}
\label{sec:ctrt_version}
By default, after computing the analysis graph, \plai produces an
output that is easy for the programmer to inspect, \emph{i.e.}, close
to the source program, where the abstract semantics of each predicate
is the \emph{combination} of all its versions.
That is, for a predicate \pred, the multiple nodes
\(\{%
  \tuple{\lambda^c_1,\pred,\lambda^s_1},%
  \ldots,%
  \tuple{\lambda^c_n,\pred,\lambda^s_n}%
\}\)
in the analysis graph corresponding to the different calling modes of
\pred, are \emph{collapsed} into a single node
\(\tuple{%
  \sqcup\{\lambda^c_1,\ldots,\lambda^c_n\},%
  \pred,%
  \sqcup\{\lambda^s_1,\ldots,\lambda^s_n\}%
}.\)
This approach, while convenient for the end-user, may incur a
significant loss of precision,
not due to the analysis itself, but rather to the over-simplification
of the abstractions.
To address this issue, \plai is also capable of producing an output
where multiple versions of each original predicate \pred, \emph{i.e.},
multiple nodes for \pred, are \emph{materialized} into different
predicates 
which, through a program transformation, are ``folded back'' to the
program.
That is, for a predicate \pred, the multiple nodes:
\(\{%
  \tuple{\lambda^c_1,\pred,\lambda^s_1},%
  \ldots,%
  \tuple{\lambda^c_n,\pred,\lambda^s_n}%
\}\)
in the analysis graph corresponding to the different calling modes of
\pred{} are \emph{materialized} into different predicates for each version:
\(\{%
  \tuple{\lambda^c_1,\pred_1,\lambda^s_1},%
  \ldots,%
  \tuple{\lambda^c_n,\pred_n,\lambda^s_n}%
\}.\)

\begin{algorithm}[t]
  \caption{$\mathsf{vers}$ program transformation.}
  \label{alg:vers}
  \scriptsize
  \begin{algorithmic}[1]
    \algrenewcommand\alglinenumber[1]{\color{gray}\scriptsize #1}
  \hspace*{-2.5em}
    \begin{minipage}[t]{.57\textwidth}
      \Function{$\mathsf{vers}$}{}%
        \State $V \gets \emptyset$
        \For{$\pred$ in the program $\prog$}
          \State $V \gets \mathsf{dump}(\pred,V)$\label{alg:vers:vers:dump}
          \State $\prog \gets \prog \setminus \setcomp{\clause \in \prog}{\head \textit{ unifies with } \pred}$
          \State $\A \gets \A \setminus \{\acCall{\pred}{\_} \in \A\}$
          \State $\A \gets \A \setminus \{\acSucc{\pred}{\_}{\_} \in \A\}$
        \EndFor
        \For{$\clause \in \prog$}
          \State $\body' \gets \mathsf{replace}(\body,V)$\label{alg:vers:vers:replace}
          \State $\prog \gets \prog \cup \{\head \texttt{ :- } \body'\}$\label{alg:vers:vers:newClause}
        \EndFor
      \EndFunction
      \State\
      \Function{$\mathsf{dump}$}{\head,$V$}
        \State $R \gets \{\clause \in \prog\}$
        \For{$\tuple{\lambda^c,\head,\lambda^s} \in \anGraph$}\label{alg:vers:dump:for}
          \State $\head' \gets \mathsf{newHead}(\head)$\label{alg:vers:dump:newHead}
          \State $R' \gets \setcomp{\head'~\texttt{:- } \body}{\clause \in R}$\label{alg:vers:dump:rules}
          \State $\prog \gets \prog \cup R'$
          \State $\A \gets \A \cup \setcomp{\acCall{\head'}{F_c}}{\acCall{\head}{F_c} \in \A}$\label{alg:vers:dump:call}
          \State $\A \gets \A \cup \setcomp{\acSucc{\head'}{F_c}{F_s}}{\acSucc{\head}{F_c}{F_s} \in \A}$\label{alg:vers:dump:succ}
          \State $\anGraph \gets \anGraph \cup \{\tuple{\lambda^c,\head',\lambda^s}\}$\label{alg:vers:dump:graph}
          \State $V \gets V \cup \{\head:\head'\}$\label{alg:vers:dump:map}
        \EndFor
        \State \Return $V$
      \EndFunction
    \end{minipage}
    \hspace*{2.5em}
    \begin{minipage}[t]{.39\textwidth}
      \Function{$\mathsf{replace}$}{\body,$V$}
        \If{$\body = \emptyGoal$}
          \State \Return $\emptyGoal$
        \Else\ $\body = L :: G$
          \If{$\exists H~s.t.~\exists! \tuple{\lambda^c,L,\lambda^s} \in \anGraph$
              \\[.2mm]\hspace*{5.9em}$s.t.~\exists! \tuple{\lambda^c,\head,\lambda^s} \in \anGraph$
              \\[.2mm]\hspace*{5.9em}$s.t.~L : \head \in V$}\label{alg:vers:replace:rep}
            \State \Return $H :: \mathsf{replace}(G,V)$
          \Else
            \State \Return $L :: \mathsf{replace}(G,V)$
          \EndIf
        \EndIf
      \EndFunction
    \end{minipage}
  \end{algorithmic}
\end{algorithm}

A high-level view of the program transformation is depicted in
\cref{alg:vers}.
For each predicate in the program, we first dump its versions
(line~\ref{alg:vers:vers:dump}).
That is, for each version of a predicate in the analysis graph
(line~\ref{alg:vers:dump:for}), a copy is created by generating a new
head (line~\ref{alg:vers:dump:newHead}) and attaching to it the
corresponding rules (line~\ref{alg:vers:dump:rules}), assertions
(lines~\ref{alg:vers:dump:call}--\ref{alg:vers:dump:succ}), and
analysis graph node (line~\ref{alg:vers:dump:graph}), updating the
program, assertion set, and analysis graph accordingly.
Additionally, a \emph{versions mapping} $V$ is created in order to
record which versions correspond to the original predicates
(line~\ref{alg:vers:dump:map}).
Then, each rule in the program is rewritten by replacing calls to the
original predicates, under a specific calling mode, with calls to the
corresponding specialized predicates generated by $\mathsf{dump}$
(lines~\ref{alg:vers:vers:replace} and~\ref{alg:vers:vers:newClause}).
Concretely, during the traversal of a rule's body, any literal
corresponding to a predicate call for which a specialized version
exists for the given call pattern is replaced by a call to that
specialized version (line~\ref{alg:vers:replace:rep}).

\begin{example}[Exploiting versions]\itshape
  Applying the $\mathsf{vers}$ transformation to the program in
  \cref{code:member} (with the analysis information of the \etermsexp
  abstract domain~\citep{eterms-sas02})
  generates two
  versions of the predicate for its
  possible abstract call patterns ($\mathit{member\_1/2}$ and $\mathit{member\_2/2}$):
  {\rm\begin{lstlisting}
(*\bfseries member*)(X,tree(_,L,_)) :- member_1(X,L). (*~~~~~~~~~~~~~~*)(*\bfseries member*)(X,[X|L]) :- list(num,L).
(*\bfseries member*)(X,tree(_,_,R)) :- member_1(X,R). (*~~~~~~~~~~~~~~*)(*\bfseries member*)(X,[_|L]) :- member_2(X,L).
(*\bfseries member*)(X,tree(X,L,R)) :- tree(num,L), tree(num,R).
  \end{lstlisting}}

  For instance, for the analysis version with calling abstraction
  $\{X/\mathit{term},\allowbreak L/\mathit{list}(\mathit{num})\}$, the $\mathit{member\_2/2}$
  version with success abstraction $\{X/\mathit{num}, L/\mathit{rt1}\}$ is
  created.
  {\rm\begin{lstlisting}
:- true pred member_2(X,L) : (term(X), list(num,L)) => (num(X), rt1(L)).
:- prop rt1/1. (*\bfseries rt1*)([A|B]) :- num(A), list(num,B). %
  \end{lstlisting}}
  The obtained success abstraction is \emph{strictly more precise}
  than that obtained by analyzing the program in \cref{code:member}
  with the \etermsexp abstract domain.
  Therefore, all assertions proven to be checked in the original
  program remain checked.
{\rm\begin{lstlisting}
:- checked pred member_2(X,S) : (var(X), list_or_tree(num,S)) => num(X).
\end{lstlisting}}
However, without versioning, recursive calls to $\mathit{member/2}$ must be
checked at run time. In contrast, with versions, recursive calls are directed
to the specialized predicate version, for which the assertions have already
been statically verified. As a result, additional run-time checks are unnecessary.
\end{example}
\section{Experimental evaluation}
\label{sec:experiments}
In order to assess the benefits of the proposed techniques, we
evaluate two sets of benchmarks. These benchmarks include a variety
of examples, ranging from a collection of classic benchmarks for
Prolog
to a subset of libraries of the \ciao system. While the 
predicates in the first set often have a single assertion per
predicate, those in the second set typically 
present more assertions per definition, describing
several possible call patterns.
Our objective is to measure the effects of applying different
techniques for the optimization of run-time checks. To this end, instead of
measuring the number of assertions that are completely checked, \emph{i.e.}, those
that are reduced to $true$, we count the number of properties that are
simplified within the assertions. We use this metric
because assertions can sometimes contain one or more properties that
are hard to verify, while other properties in the assertion are
verified.  The metric chosen allows us to measure the effects of
different techniques at a finer level of 
granularity. 
Then, we also evaluate the impact of these techniques on the execution
times of the programs when executed under run-time checking semantics.
The experiments were run on a MacBook Air
with the Apple M1 chip and 16 GB of RAM, with
a per-run timeout of 3 minutes.

\begin{table}
  \caption{Number of reduced properties (\%). \label{tab:props}}
  \hspace*{-6mm}
  {\tablefont\scriptsize
    \centering\makebox[\textwidth][l]{%
    \setlength{\tabcolsep}{3pt}
    \hspace*{-1.5em}
    \begin{tabular}{||p{16mm}|>{\raggedleft\arraybackslash}p{22mm}|>{\raggedleft\arraybackslash}p{22mm}|>{\raggedleft\arraybackslash}p{22mm}|>{\raggedleft\arraybackslash}p{28mm}|>{\raggedleft\arraybackslash}p{25mm}||}
     \hline \hline
    \multicolumn{1}{||c|}{\textbf{module}} &
    \multicolumn{1}{|c|}{\ct} &
    \multicolumn{1}{|c|}{\ctrt} &
    \multicolumn{1}{|c|}{\ctrtf} &
    \multicolumn{1}{|c|}{\ctrtfetsh} &
    \multicolumn{1}{|c||}{\ctrtfshet} \\ \hline \hline
    \texttt{exp}      & 18/21 (85.71\%)   & 18/21 (85.71\%)   & 18/21 (85.71\%)   & 27/30 (90.00\%)              & 18/21 (85.71\%)          \\
    \texttt{factorial}& 4/6 (66.67\%)     & 4/6 (66.67\%)     & 4/6 (66.67\%)     & 4/6 (66.67\%)                & \textbf{10/12 (83.33\%)} \\
    \texttt{fft}      & 69/95 (72.63\%)   & 92/95 (96.84\%)   & 91/95 (95.79\%)   & 175/179 (97.77\%)            & 214/218 (98.17\%)        \\
    \texttt{fib}      & 16/18 (88.89\%)   & 16/18 (88.89\%)   & 16/18 (88.89\%)   & \textbf{40/42 (95.24\%)}     & 28/30 (93.33\%)          \\
    \texttt{guardians}& 51/54 (94.44\%)   & 51/54 (94.44\%)   & 51/54 (94.44\%)   & 123/126 (97.62\%)            & 69/72 (95.83\%)          \\
    \texttt{hamming}  & 94/96 (97.92\%)   & 94/96 (97.92\%)   & 94/96 (97.92\%)   & 299/303 (98.68\%)            & 148/150 (98.67\%)        \\
    \texttt{hanoi}    & 19/24 (79.17\%)   & 19/24 (79.17\%)   & 19/24 (79.17\%)   & 37/42 (88.10\%)              & 52/57 (91.23\%)          \\
    \texttt{jugs}     & 42/45 (93.33\%)   & 42/45 (93.33\%)   & 42/45 (93.33\%)   & 66/69 (95.65\%)              & 54/57 (94.74\%)          \\
    \texttt{knights}  & 58/60 (96.67\%)   & 58/60 (96.67\%)   & 58/60 (96.67\%)   & 105/107 (98.13\%)            & 70/72 (97.22\%)          \\
    \texttt{mmatrix}  & 14/17 (82.35\%)   & 14/17 (82.35\%)   & 14/17 (82.35\%)   & 14/17 (82.35\%)              & 14/17 (82.35\%)          \\
    \texttt{nreverse} & 13/15 (86.67\%)   & 13/15 (86.67\%)   & 13/15 (86.67\%)   & 22/24 (91.67\%)              & \textbf{46/48 (95.83\%)} \\
    \texttt{poly}     & 60/63 (95.24\%)   & 60/63 (95.24\%)   & 60/63 (95.24\%)   & 159/162 (98.15\%)            & 249/252 (98.81\%)        \\
    \texttt{primes}   & 28/30 (93.33\%)   & 28/30 (93.33\%)   & 28/30 (93.33\%)   & 52/54 (96.30\%)              & 37/39 (94.87\%)          \\
    \texttt{progeom}  & 67/69 (97.10\%)   & 67/69 (97.10\%)   & 67/69 (97.10\%)   & 230/232 (99.14\%)            & 98/100 (98.00\%)         \\
    \texttt{qsort}    & 27/28 (96.43\%)   & 27/28 (96.43\%)   & 27/28 (96.43\%)   & 36/37 (97.30\%)              & 57/58 (98.28\%)          \\
    \texttt{queens}   & 24/26 (92.31\%)   & 24/26 (92.31\%)   & 24/26 (92.31\%)   & 54/56 (96.43\%)              & 41/43 (95.35\%)          \\
    \texttt{serialize}& 19/21 (90.48\%)   & 19/21 (90.48\%)   & 19/21 (90.48\%)   & 29/31 (93.55\%)              & 29/31 (93.55\%)          \\
    \texttt{tak}      & 8/12 (66.67\%)    & 8/12 (66.67\%)    & 8/12 (66.67\%)    & 8/12 (66.67\%)               & \textbf{20/24 (83.33\%)} \\
    \hline
    \textbf{total}    & 631/700 (90.14\%) & 654/700 (93.41\%) & 653/700 (93.28\%) & \textbf{1480/1529 (96.80\%)} & 1254/1301 (96.39\%)      \\
    \hline
    \texttt{atom\_concat}  & 0/5 (0.00\%)      & 0/5 (0.00\%)      & 0/5 (0.00\%)      & \textbf{13/15 (86.67\%)}     & 0/20 (0.00\%)       \\
    \texttt{formulae}      & 12/21 (57.14\%)   & 12/21 (57.14\%)   & 16/21 (76.19\%)   & 26/35 (74.29\%)              & 44/70 (62.86\%)     \\
    \texttt{iso\_char}     & 13/39 (33.33\%)   & 22/39 (56.41\%)   & 13/39 (33.33\%)   & 56/87 (64.37\%)              & 37/93 (39.78\%)     \\
    \texttt{lists}         & 51/106 (48.11\%)  & 50/106 (47.17\%)  & 51/106 (48.11\%)  & 108/183 (59.02\%)            & 88/174 (50.57\%)    \\
    \texttt{numlists}      & 6/11 (54.55\%)    & 6/11 (54.55\%)    & 6/11 (54.55\%)    & 11/16 (68.75\%)              & 11/16 (68.75\%)     \\
    \texttt{random\_utils} & 34/56 (60.71\%)   & 35/56 (62.50\%)   & 35/56 (62.50\%)   & 52/73 (71.23\%)              & 52/73 (71.23\%)     \\
    \texttt{sets}          & 0/46 (0.00\%)     & 0/46 (0.00\%)     & 0/46 (0.00\%)     & 2/49 (4.08\%)                & 0/46 (0.00\%)       \\
    \texttt{amqueue}       & 7/34 (20.59\%)    & 7/34 (20.59\%)    & 7/34 (20.59\%)    & \textbf{108/147 (73.47\%)}   & 32/123 (26.02\%)    \\
    \texttt{binary\_tree}  & 6/19 (31.58\%)    & 9/19 (47.37\%)    & 9/19 (47.37\%)    & \textbf{33/38 (86.84\%)}     & 19/38 (50.00\%)     \\
    \texttt{btree}         & timeout           & 78/113 (69.03\%)  & 68/113 (60.18\%)  & 345/426 (80.99\%)            & 467/558 (83.69\%)   \\
    \texttt{heap}          & 77/90 (85.56\%)   & 79/90 (87.78\%)   & 85/90 (94.44\%)   & 187/192 (97.40\%)            & 451/456 (98.90\%)   \\
    \texttt{rbtree}        & 101/132 (76.52\%) & 122/132 (92.42\%) & 124/132 (93.94\%) & 112/120 (93.33\%)            & 112/120 (93.33\%)   \\
    \texttt{sets\_ops}     & 32/77 (41.56\%)   & 32/77 (41.56\%)   & 32/77 (41.56\%)   & 103/161 (63.98\%)            & 112/158 (70.89\%)   \\
    \hline
    \textbf{total}         & 339/749 (45.26\%) & 382/749 (51.00\%) & 446/749 (59.55\%) & \textbf{1156/1542 (74.97\%)} & 1425/1945 (73.26\%) \\
    \hline\hline
    \end{tabular}}%
  }
\end{table}
Table~\ref{tab:props}
presents the results of the first evaluation.
Each column displays the number of properties actually verified (``checked'')
at compile time, the total number of properties in the corresponding module, and
the percentage of properties checked.
These experiments use the \etermsexp and \shfrexp abstract domains.
Column \ct presents results when using the information obtained from 
the classic fixpoint with default
semantics. Column \ctrtf shows results when executing the fixpoint
with run-time checking semantics, as defined in
Section~\ref{sec:assrt:rt}.
Column \ctrtfetsh (resp.~\ctrtfshet) displays results when executing
the fixpoint with run-time checking semantics, followed by the versions
transformation (\emph{i.e.}, materializing the versions contained 
in the multivariant information of \etermsexp), and then 
analyzing these using \shfrexp (resp.\ with the multivariant information of
\shfrexp and analyzing using \etermsexp).
A domain combination approach can be used to join versions
but we wanted to observe 
the effects of the versions of each domain separately.
The results show that 
applying run-time checking semantics to the analyzer increases the number
of verified properties. The transformation obviously raises the total
property count (since assertions are duplicated across versions), but
the usage of multivariant information yields overall better
percentages. Notably, different modules show varied improvements based
on whether the multivariant information inferred from \shfrexp or
\etermsexp is used for the versions transformation.%

\begin{table}
  \caption{Execution time, ms (speedup \emph{w.r.t.} \textbf{rt}). \label{tab:times}}
  \hspace*{-4mm}
  {\tablefont\scriptsize
    \centering\makebox[\textwidth][l]{%
    \setlength{\tabcolsep}{3pt}
    \hspace*{-1.5em}
    \begin{tabular}{||p{16mm}|>{\raggedleft\arraybackslash}p{11mm}|>{\raggedleft\arraybackslash}p{20mm}|>{\raggedleft\arraybackslash}p{20mm}|>{\raggedleft\arraybackslash}p{20mm}|>{\raggedleft\arraybackslash}p{19.5mm}|>{\raggedleft\arraybackslash}p{19.5mm}||}
     \hline \hline
    \multicolumn{1}{||c|}{\textbf{module}} &
    \multicolumn{1}{|c|}{\textbf{rt}} &
    \multicolumn{1}{|c|}{\ct} &
    \multicolumn{1}{|c|}{\ctrt} &
    \multicolumn{1}{|c|}{\ctrtf} &
    \multicolumn{1}{|c|}{\ctrtfetsh} &
    \multicolumn{1}{|c||}{\ctrtfshet} \\ \hline \hline
    \texttt{exp}        &    38.093 &      5.419 (7$\times$) &      5.454 (7$\times$) &      5.419 (7$\times$) &      5.422 (7$\times$)  &             5.421 (7$\times$) \\
    \texttt{factorial}  &     0.357 &      0.183 (2$\times$) &      0.183 (2$\times$) &      0.182 (2$\times$) &      0.181 (2$\times$)  &            0.027 (13$\times$) \\
    \texttt{fft}        & 15,057.10 & 11,468.4 (1.3$\times$) &   39.954 (376$\times$) &  1,303.05 (11$\times$) &  1,285.54 (12$\times$)  &         1,281.08 (12$\times$) \\
    \texttt{fib}        &     6.871 &    0.043 (159$\times$) &    0.048 (143$\times$) &    0.043 (159$\times$) &    0.043 (159$\times$)  &           0.043 (159$\times$) \\
    \texttt{guardians}  &  3,308.72 &   2.084 (1587$\times$) &   2.222 (1489$\times$) &   2.082 (1589$\times$) &   2.075 (1594$\times$)  &          2.075 (1594$\times$) \\
    \texttt{hamming}    &  4,098.18 &    8.210 (499$\times$) &    9.274 (441$\times$) &    8.298 (493$\times$) &    8.166 (501$\times$)  &           8.123 (504$\times$) \\
    \texttt{hanoi}      &    64.731 &     1.462 (44$\times$) &     1.476 (43$\times$) &     1.462 (44$\times$) &     1.462 (44$\times$)  & \textbf{0.045 (1438$\times$)} \\
    \texttt{jugs}       &     0.712 &     0.012 (59$\times$) &     0.012 (59$\times$) &     0.012 (59$\times$) &     0.012 (59$\times$)  &            0.012 (59$\times$) \\
    \texttt{knights}    &  8,440.13 &   106.201 (79$\times$) &   110.018 (76$\times$) &   111.269 (75$\times$) &   107.094 (78$\times$)  &          105.934 (79$\times$) \\
    \texttt{mmatrix}    &     1.471 &     0.092 (16$\times$) &     0.093 (16$\times$) &     0.092 (16$\times$) &     0.092 (16$\times$)  &            0.092 (16$\times$) \\
    \texttt{nreverse}   &  5,020.80 &   10.800 (464$\times$) &   11.179 (449$\times$) &   10.793 (465$\times$) &   10.786 (465$\times$)  & \textbf{1.141 (4400$\times$)} \\
    \texttt{poly}       &   192.200 &    0.767 (251$\times$) &    0.831 (231$\times$) &    0.766 (250$\times$) &    0.736 (261$\times$)  &           0.736 (261$\times$) \\
    \texttt{primes}     &     5.269 &    0.015 (351$\times$) &    0.017 (309$\times$) &    0.015 (351$\times$) &    0.015 (351$\times$)  &           0.015 (351$\times$) \\
    \texttt{progeom}    &  204,112 &   727.256 (280$\times$) &  793.123 (257$\times$) &  727.265 (280$\times$) &  726.545 (280$\times$)  &         727.287 (280$\times$) \\
    \texttt{qsort}      &    18.089 &     0.369 (49$\times$) &     0.379 (48$\times$) &     0.380 (48$\times$) &     0.370 (49$\times$)  &  \textbf{0.062 (292$\times$)} \\
    \texttt{queens}     &   326,621 & 1,358.14 (240$\times$) & 1,515.82 (215$\times$) & 1,359.46 (240$\times$) & 1,359.02 (240$\times$)  &        1,359.40 (240$\times$) \\
    \texttt{serialize}  &     2.056 &     0.021 (98$\times$) &     0.022 (93$\times$) &     0.021 (98$\times$) &     0.021 (98$\times$)  &            0.021 (98$\times$) \\
    \texttt{tak}        &   436.523 &    213.487 (2$\times$) &    214.284 (2$\times$) &    213.103 (2$\times$) &     213.183 (2$\times$) &  \textbf{1.824 (239$\times$)} \\
    \hline
    \textbf{geom.~mean} &   184.282 &     2.953 (62$\times$) &     2.263 (81$\times$) &     2.628 (70$\times$) &      2.607 (71$\times$) &  \textbf{1.185 (156$\times$)} \\
    \hline
    \texttt{atom\_concat} &   14.052 &   14.068 (1$\times$) &   14.886 (1$\times$) &   14.071 (1$\times$) &            7.044 (2$\times$) &         18.153 (1$\times$) \\
    \texttt{formulae}     & 1,365.66 & 0.375 (3641$\times$) & 0.377 (3622$\times$) & 0.376 (3633$\times$) &         0.376 (3633$\times$) &       0.378 (3613$\times$) \\
    \texttt{iso\_char}    &    5.965 &    5.967 (1$\times$) &    3.476 (2$\times$) &    5.965 (1$\times$) &          2.293 (2.5$\times$) &          5.523 (1$\times$) \\
    \texttt{lists}        &   730.89 &  21.678 (34$\times$) &  25.239 (29$\times$) &  21.690 (34$\times$) & \textbf{18.402 (40$\times$)} &        21.472 (34$\times$) \\
    \texttt{numlists}     &  390.163 &  229.500 (2$\times$) &  229.387 (2$\times$) &  229.392 (2$\times$) &          230.229 (2$\times$) &        230.589 (2$\times$) \\
    \texttt{sets}         &  271.956 &  271.861 (1$\times$) &  271.289 (1$\times$) &  271.861 (1$\times$) &          265.793 (1$\times$) &        267.915 (1$\times$) \\
    \texttt{binary\_tree} &    0.033 &    0.033 (1$\times$) &  0.045 (0.7$\times$) &    0.033 (1$\times$) &            0.015 (2$\times$) &          0.033 (1$\times$) \\
    \texttt{btree}        &    0.248 &              timeout &    0.149 (2$\times$) &    0.208 (1$\times$) &            0.135 (2$\times$) &          0.129 (2$\times$) \\
    \texttt{heap}         &    0.287 &    0.089 (3$\times$) &    0.084 (3$\times$) &    0.038 (8$\times$) &            0.038 (8$\times$) &          0.038 (8$\times$) \\
    \texttt{rbtree}       &    0.274 &    0.119 (2$\times$) &    0.030 (9$\times$) &    0.031 (9$\times$) &            0.031 (9$\times$) & \textbf{0.031 (9$\times$)} \\
    \texttt{sets\_ops}    &    0.676 &    0.238 (3$\times$) &    0.226 (3$\times$) &    0.228 (3$\times$) &            0.198 (3$\times$) &          0.171 (4$\times$) \\
    \hline
    \textbf{geom.~mean}   &    7.439 &    2.264 (3$\times$) &    1.542 (5$\times$) &    1.487 (5$\times$) &   \textbf{1.112 (7$\times$)} &          1.408 (5$\times$) \\
    \hline\hline
    \end{tabular}}%
  }
\end{table}
\Cref{tab:times} presents the execution times for the programs,
running with run-time checks activated, for the different
techniques. As a baseline, column \textbf{rt} shows the
execution times
without any optimization. We also show in parenthesis the
speedups obtained with respect to our baseline.

The results indicate that applying any of
the proposed techniques significantly improves program execution
times.
The fixpoint analysis with run-time semantics yields results similar
to those obtained with the transformation technique proposed by
\citet{optchk-journal-scp-short}. %
In a few cases, however, the \ctrt technique
outperforms our fixpoint implementation (\emph{e.g.}, \texttt{fft}). This
difference is caused by the widening operators in the \etermsexp
abstract domain being overly conservative very soon in the fixpoint
iteration. In that example, only one property was not checked under
the \ctrt transformation, yet this caused the fixpoint approach to be
32 times slower. Even so, it remained 12 times faster than the
baseline.
This example highlights the importance of discharging as many properties
as possible at compile-time.
It also raises questions about whether additional improvements, or new
widening operators are needed, and whether more precise operations
could further improve execution times.
Exploiting the multivariant information to \emph{automatically}
generate program versions allows speed-ups of up to 118× compared to
the \ctrt transformation (\emph{e.g.}, \texttt{tak}). In some cases, this
corresponds to a total speed-up of up to 240× with respect to the
baseline (\textbf{rt}) execution. Unlike \ctrt, which requires
manually crafting specialized versions, the new technique
materializes versions systematically. This is especially relevant in
more realistic modules, where multiple assertions and call patterns
would make a manual, per-predicate specialization impractical.
For programs where more moderate speed-ups were observed (ranging
from 2× to 10×), execution times were comparable to the more precise
approach. In all cases, both techniques were still several times, or
even orders of magnitude, faster than the baseline execution.
We also observe that the order in which the information of the
versions from different domains is applied matters, as
expected (\emph{e.g.}, \texttt{factorial}, \texttt{nreverse},
\texttt{qsort}, and \texttt{tak}, vs., \emph{e.g.},
\texttt{atom\_concat}, \texttt{iso\_char}, or \texttt{lists}). In
practice we would thus use a combined domain approach, or simply try
both orders and keep the best.

\begin{table}
  \caption{Number of reduced properties (\%) for \lpdoc. \label{tab:props_lpdoc}}
  \hspace*{-5mm}
  {\tablefont\scriptsize
    \centering\makebox[\textwidth][l]{%
    \setlength{\tabcolsep}{3pt}
    \hspace*{-1.5em}
    \begin{tabular}{||p{16mm}|>{\raggedleft\arraybackslash}p{22mm}|>{\raggedleft\arraybackslash}p{22mm}|>{\raggedleft\arraybackslash}p{22mm}|>{\raggedleft\arraybackslash}p{25mm}|>{\raggedleft\arraybackslash}p{25mm}||}
     \hline \hline
    \multicolumn{1}{||c|}{\textbf{module}} &
    \multicolumn{1}{|c|}{\ct} &
    \multicolumn{1}{|c|}{\ctrt} &
    \multicolumn{1}{|c|}{\ctrtf} &
    \multicolumn{1}{|c|}{\ctrtfetsh} &
    \multicolumn{1}{|c||}{\ctrtfshet} \\ \hline \hline
    \texttt{doctree}        & 16/48 (33.33\%)  & 19/48 (39.58\%)   & 22/48 (45.83\%)   & \textbf{62/90 (68.89\%)}   & \textbf{83/117 (70.94\%)}  \\
    \texttt{filesystem}     & 16/32 (50.00\%)  & 19/32 (59.38\%)   & 21/32 (65.62\%)   & 21/32 (65.62\%)            & 21/32 (65.62\%)            \\
    \texttt{html}           & 6/22 (27.27\%)   & 14/22 (63.64\%)   & 21/22 (95.45\%)   & 21/22 (95.45\%)            & 21/22 (95.45\%)            \\
    \texttt{html\_assets}   & 1/1 (100.00\%)   & 1/1 (100.00\%)    & 1/1 (100.00\%)    & 1/1 (100.00\%)             & 1/1 (100.00\%)             \\
    \texttt{html\_template} & 4/12 (33.33\%)   & 6/12 (50.00\%)    & 6/12 (50.00\%)    & 6/12 (50.00\%)             & 6/12 (50.00\%)             \\
    \texttt{images}         & 5/12 (41.67\%)   & 5/12 (41.67\%)    & 5/12 (41.67\%)    & 5/12 (41.67\%)             & 5/12 (41.67\%)             \\
    \texttt{index}          & 5/5 (100.00\%)   & 5/5 (100.00\%)    & 5/5 (100.00\%)    & 5/5 (100.00\%)             & 5/5 (100.00\%)             \\
    \texttt{lpdoc\_help}    & 0/2 (0.00\%)     & 0/2 (0.00\%)      & 0/2 (0.00\%)      & 0/2 (0.00\%)               & 0/2 (0.00\%)               \\
    \texttt{man}            & 0/4 (0.00\%)     & 0/4 (0.00\%)      & 4/4 (100.00\%)    & 4/4 (100.00\%)             & 4/4 (100.00\%)             \\
    \texttt{nil}            & 0/4 (0.00\%)     & 0/4 (0.00\%)      & 4/4 (100.00\%)    & 4/4 (100.00\%)             & 4/4 (100.00\%)             \\
    \texttt{refsdb}         & 17/29 (58.62\%)  & 18/29 (62.07\%)   & 18/29 (62.07\%)   & 18/29 (62.07\%)            & \textbf{27/38 (71.05\%)}   \\
    \texttt{single\_mod}    & 0/8 (0.00\%)     & 0/8 (0.00\%)      & 0/8 (0.00\%)      & 0/8 (0.00\%)               & 0/8 (0.00\%)               \\
    \texttt{texinfo}        & 12/16 (75.00\%)  & 13/16 (81.25\%)   & 16/16 (100.00\%)  & 16/16 (100.00\%)           & 16/16 (100.00\%)           \\
    \hline
    \textbf{total}          & 82/195 (42.05\%) & 100/195 (51.28\%) & 123/195 (63.08\%) & \textbf{163/237 (68.78\%)} & \textbf{193/273 (70.70\%)} \\
    \hline\hline
    \end{tabular}}%
  }
\end{table}
As a final experiment, we tested the approach on a more complex
application: the \lpdoc
documenter~\citep{lpdoc-cl2000,lpdoc-reference-manual-3.0-short}.
\lpdoc is a documentation generator for LP systems, used in both \ciao and
XSB. In addition to generating all the Ciao system manuals and XSB
manuals, it is used to generate many web-sites, class slides and
exercises, user program documentation, etc.  It comprises
approximately 22K lines of Prolog code, plus the use of many \ciao
libraries. The results on property reduction are shown in
\cref{tab:props_lpdoc}. Perhaps the most notable (and
encouraging) conclusion from this experiment is that these results do
not differ qualitatively from those on the smaller benchmarks, and
similar positive overall patterns are observed.
Overall, the \ctrtfshet analysis verified 193 properties out of a total of
273, \emph{i.e.}, 70\% of all the properties appearing in the assertions in
the code. Furthermore, the verification process detected some errors
in \lpdoc that had gone unnoticed for many years.

\section{Related approaches}
The pattern of folding assertion preconditions into a multivariant
abstract state and discharging postconditions per clause appears, at
least partially, in other approaches. Perhaps the closest to the
CiaoPP model~\citep{prog-glob-an-short} is 
soft contract verification in higher-order functional
languages~%
\citep{DBLP:conf/icfp/NguyenTH14,DBLP:conf/popl/NguyenGTH18}, as well as
the gradual-typing-optimization line~%
\citep{DBLP:conf/popl/TakikawaFGNVF16-short,DBLP:conf/popl/RastogiSFBV15-short},
which has also aimed in part at generating residual programs with
optimized run-time checks. 
Also related, but not necessarily aimed at producing optimized programs,
are assume-based imperative abstract interpreters such as
\textsc{Astr\'ee}~\citep{astree-esop05},
Frama-C/EVA~\citep{DBLP:journals/fac/KirchnerKPSY15}, and
Goblint~\citep{DBLP:conf/sas/SchwarzSAEHSV21}. 
Deductive Verification Condition (VC)-based verifiers
(\emph{e.g.}, Dafny~\citep{DBLP:conf/lpar/Leino10},
Why3~\citep{DBLP:conf/esop/FilliatreP13},
F$^{\star}$~\citep{DBLP:conf/popl/SwamyHKRDFBFKKZ16}) can in
principle discharge sub-clauses via independent VCs, but differ in several
respects: their objective is a binary verification verdict, 
they do not produce optimized programs with simplified
checks, and they inherit the solver's incompleteness on recursive and
quantified goals. In contrast, our approach based on abstract interpretation
always terminates (guaranteed by widening)
and unverified
properties are turned into run-time checks rather than
blocking deployment.
In this context, our contribution is to integrate the run-time checking
semantics directly into the multivariant top-down
fixpoint via the alternative call and success abstractions
(\cref{alg:abscalls,alg:absprimes}), avoiding the auxiliary program
transformation of~\cite{optchk-journal-scp-short}; to exploit the inferred
multivariant information to materialize specialized
predicate versions (Algorithm~\ref{alg:vers}); and to provide
performance results for this approach.

\section{Conclusions and future work}
\label{sec:conclusions}
We have addressed the problem of reducing run-time overhead in the
context of verification frameworks that combine compile-time and
run-time checking of user-provided assertions.
To this end, we have described how the multivariant, top-down abstract
interpretation algorithm can be adapted to more precisely analyze
programs that are executed under optional run-time assertion semantics.
We have studied how the multivariant information abstracted by
these frameworks can be exploited to increase the number of program
properties that are checked at compile time. 
Our experimental results show
that these techniques are not only effective in increasing the
number of properties that are checked at compile time,
but also in reducing execution times
under run-time assertion checking semantics.
We presented our approach in the context of the \ciao Prolog language,
but the \ciao approach to combining static and dynamic analysis is
general and system-independent, as well as the techniques used herein,
so we expect the results to carry over to other (dynamic) logic
languages.
Future work includes investigating further the impact of abstract
specialization, and, as mentioned previously, studying the effects of
other domains and domain combinations. 

\ \\
\noindent
\textbf{Competing interests:}
The authors declare none.
\bibliographystyle{tlplike}

\end{document}